\documentclass[journal]{IEEEtran}
\ifCLASSINFOpdf
  \usepackage[pdftex]{graphicx}
  \graphicspath{{../pdf/}{../jpeg/}}
  \DeclareGraphicsExtensions{.pdf,.jpeg,.png}
\else
  \usepackage[dvips]{graphicx}
  \graphicspath{{../eps/}}
  \DeclareGraphicsExtensions{.eps}
\fi
\usepackage{amsfonts}

\usepackage{multirow}

\begin{document}
%
\title{Truncated Analytic Moment Analysis and Its Hybrid-Field Contrast in Grating-based X-ray Phase Contrast Imaging}
%
%
%

\author{Chengpeng~Wu, Li~Zhang, Xinbin~Li, Zhiqiang~Chen, Yuxiang~Xing, Xiaohua~Zhu, Hewei~Gao
\thanks{Manuscript received December 1, 2019. This work was supported in part by the National Natural Science Foundation of China (No. 11435007 and No. 61527807), National Key R\&D Program of China (2016YFB1101203) and the New Faculty Startup Fund from Tsinghua University, Beijing, China. (Corresponding authors: Hewei Gao, Li Zhang).}
\thanks{C. Wu, L. Zhang, X. Li, Z. Chen, Y. Xing, X. Zhu and H. Gao are with the Department of Engineering Physics, Tsinghua University, Beijing, China, and also with the Key Laboratory of Particle and Radiation Imaging (Tsinghua University), Ministry of Education, China (e-mail: hwgao@tsinghua.edu.cn, zli@mail.tsinghua.edu.cn).}}

\markboth{Wu \MakeLowercase{\textit{et al.}}: Truncated Analytic Moment Analysis and Its Hybrid-Field Contrast in GPCI}{}
%



\maketitle

\begin{abstract}
For grating-based x-ray phase contrast imaging (GPCI), a multi-order moment analysis (MMA) has been recently developed to obtain multiple contrasts from the ultra-small-angle x-ray scattering distribution, as a novel  information retrieval approach that is totally different from the conventional Fourier components analysis (FCA). In this paper, we present an analytic form of MMA in theory that can retrieve multiple contrasts directly from raw phase-stepping images, with no scattering distribution involved. For practical implementation, a truncated analytic analysis (called as TA-MMA) is adopted and it is hundreds of times faster in computation than the original deconvolution-based MMA (called as DB-MMA). More importantly, TA-MMA is proved to establish a quantitative connection between FCA and MMA, i.e., the first-order moment computed by TA-MMA is essentially the product of the phase contrast and the dark-field contrast retrieved by FCA, providing a new physical parameter for GPCI. The new physical parameter, in fact, can be treated as a ``hybrid-field'' contrast as it fuses the original phase contrast and dark-field contrast in a straightforward manner, which may be the first physical fusion contrast in GPCI to our knowledge and may have a potential to be directly used in practical applications.
\end{abstract}


%
\IEEEpeerreviewmaketitle

\section{Introduction}
\bstctlcite{myref:BSTcontrol}
%
%
%
%
\IEEEPARstart{G}{rating-based} x-ray phase contrast imaging (GPCI) has been developed as a promising technology in x-ray phase contrast imaging field. Considering that the traditional x-ray imaging is based on the attenuation index, which is three orders of magnitude lower than the refractive index decrement for materials comprised of light (low-Z) elements like soft tissues under the low energy range of x-ray (15-60 keV)  \cite{momose2002phase}, GPCI utilizing the phase shift as the contrast has been proposed and developed rapidly in the last decade. Comparing with other x-ray phase contrast imaging methods such as crystal interferometry  \cite{bonse1965x,momose1995phase,momose1996phase}, analyzer-based imaging  \cite{davis1995phase,ingal1995x} and the propagation-based imaging  \cite{wilkins1996phase,cloetens2006quantitative}, GPCI  \cite{pfeiffer2006phase,pfeiffer2008hard} can overcome the strict limitations by the coherence of x-ray sources and therefore can be implemented using a conventional x-ray source nowadays, which makes it possible for wide applications. In addition to the absorption contrast (ATC) and the differential phase contrast (DPC), GPCI can obtain the dark-field contrast (DFC) simultaneously  \cite{pfeiffer2008hard,pfeiffer2012milestones}, which delivers the ultra-small-angle x-ray scattering (USAXS) information of the sample on sub-pixel scales. Recently, GPCI shows great potentials for clinical diagnosis such as lung imaging  \cite{schleede2012emphysema,meinel2014lung,gradl2018dynamic} and breast imaging  \cite{anton2013grating, li2018diagnosis, baran2018high}.

\begin{figure}[!t]
\centering
\includegraphics[width=0.48\textwidth]{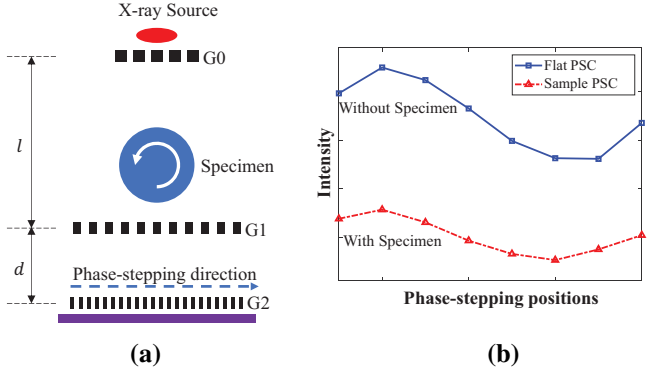}\\%
\textbf{(a)} \qquad \qquad \qquad \qquad \qquad \qquad \textbf{(b)}
\caption{The schematic diagram of  three-gratings GPCI systems like the Talbot-Lau interferometry and the geometrical-projection system (a) and its phase-stepping curves acquired with (the sample PSC) and without (the flat PSC) the specimen in a pixel (b).}
\label{fig:system}
\end{figure}

From the perspective of imaging principles, GPCI systems can be categorized into two types: the coherent systems (such as the Talbot \cite{david2002differential,momose2003demonstration} and Talbot-Lau interferometry \cite{pfeiffer2006phase}), and the incoherent systems (such as the geometrical-projection system \cite{huang2009alternative}). The principle of all GPCI systems is to measure the subtle difference due to the refraction and scattering caused by the sample. In practical imaging configuration of all GPCI systems, the phase-stepping strategy is the most commonly used approach to acquire the periodic pattern of the intensity signal at each pixel. As illustrated in Fig. \ref{fig:system}, in the phase-stepping approach, one of the gratings (usually the last grating G2) is moved along the transverse direction perpendicular to the grating lines step by step over one grating period and the detector acquires an image at each step to obtain a so-called phase-stepping curve (PSC) at each pixel. After two PSCs with and without the sample object are acquired, the multiple contrasts such as ATC, DPC and DFC can be retrieved using an information retrieval algorithm, which is a critical data processing and has been actively investigated in the field of GPCI. 

In medical applications, it is of utmost importance to present to radiologists all clinically relevant information in as compact a way as possible. Hence, the need arises for a method to combine two or more of the above mentioned contrasts in GPCI into one image containing best information relevant for diagnosis. Until now, contrast fusion methods  in GPCI are post-processing after obtaining multiple contrasts, either in image domain  \cite{wang2013image,scholkmann2014new} or fourier domain  \cite{coello2017fourier}, and are separated from the information retrieval processing.

So far, there are mainly two types of information retrieval methods in GPCI. When the Talbot interferometry was presented in 2003, Momose et al. assumed that the moir\'e fringes are approximately cosinusoidal  \cite{momose2003demonstration}. After the phase-stepping method was introduced into GPCI  \cite{weitkamp2005x,momose2005recent}, PSCs are taken as a pseudo-cosinusoidal curve  \cite{pfeiffer2006phase,pfeiffer2008hard} and regular contrasts are expressed directly by parameters of PSCs. This information retrieval method is commonly used in GPCI and is referred as Fourier components analysis (FCA) in this paper. However, FCA has an inherent phase-wrapping problem, which could be a serious problem for practical applications \cite{jerjen2011reduction,epple2014phase,rodgers2018quantitative}. 

Recently, Modregger et al. developed a new type of information retrieval method, in which one first obtains the underlying USAXS distribution by the Lucy-Richardson deconvolution and then computes the multiple contrasts as moments of the USAXS distribution \cite{modregger2012imaging}. It is referred as deconvolution-based MMA or DB-MMA in this paper, whose physical model of the three modalities was developed in Ref.  \cite{khelashvili2005physical}. The essential foundation of DB-MMA is the well-established convolution relationship that the sample PSC can be considered as the convolution of the flat PSC (without sample object) with the USAXS distribution  \cite{wang2009quantitative}. Compared with FCA, DB-MMA can provide contrasts with relatively lower noise  \cite{weber2013increasing}, naturally free from the phase-wrapping problem while preserving sufficient structural details. In addition, DB-MMA extends the possible complementary contrasts greatly besides the regular ones by computing higher order moments. However, the time-consuming deconvolution process and the difficulty in determining a proper stopping criteria for the iteration make the original DB-MMA less-favorable for time-sensitive applications. 

To overcome these limitations of FCA and DB-MMA, we therefore propose an analytic multi-order moment analysis in theory that can obtain regular contrasts directly from PSCs with no extra processes like the deconvolution. For practical implementation, a truncated analytic analysis (called as TA-MMA) is adopted and it's demonstrated that TA-MMA is efficient and stable, leading to hundreds of times faster in computation than the original DB-MMA. Under a similar purpose, Li et al.  \cite{li2019comparative} introduced the principal component analysis (PCA) to simplify the iteration process of deconvolution, which is also of much higher efficiency than DB-MMA. However, the PCA-based analysis has no physical interpretations and only considers first three principal singular values, leading to partial loss of the information. With a special cosine expression of the USAXS distribution satisfing both the cosine model of PSCs and the convolution relationship, Li et al. revealed an intrinsic relationship between FCA and DB-MMA, and found the phase contrast and dark-field contrasts computed by DB-MMA are dependent with each other  \cite{li2019comparative}. Besides, it's noticed that recently another independent work by Modregger et al. \cite{modregger2018direct} also proposes a direct MMA form (called as D-MMA) for edge-illumination imaging, which provides almost equivalent contrasts with DB-MMA while speeding up data analysis by almost three orders of magnitude. The foundation of D-MMA is also the convolution relationship between the PSCs and the USAXS distribution, while it is established under the assumption that original circular convolution formula in DB-MMA can be approximated with the linear convolution formula in D-MMA. Unfortunately, it is not directly applicable to GPCI, whose PSCs have the cosine nature, not satisfying the linear convolution assumption in D-MMA. As demonstrated by Zhu et. al. \cite{zhu2019direct}, D-MMA can be extended to GPCI after some data pre-processing and it can be used under some conditions.

More importantly, in this paper, it is proved that TA-MMA establishs a quantitative connection between FCA and MMA, i.e., the first-order moment computed by TA-MMA can be considered as the product of the phase contrast and the dark-field contrast retrieved by FCA, providing a new physical parameter for GPCI. The new physical parameter fuses the original phase contrast and dark-field contrast in a straightforward manner and therefore can be treated as a ``hybrid-field'' contrast, which may be the first physical fusion contrast in GPCI to our knowledge and has a potential to be directly used in practical applications. 
\section{\label{sec:methods}Methods}
\subsection{Two main information retrieval approaches: FCA and DB-MMA}
In GPCI, there are mainly two types of information retrieval approaches, i.e. FCA and DB-MMA. The FCA method is based on the cosine assumption for both the flat PSC and the sample PSC   \cite{pfeiffer2006phase,pfeiffer2008hard,weitkamp2005x}, i.e.,
\begin{eqnarray}
\label{equ:1} &f(\phi)=a_0^f+a_1^f\cos(\phi+\varphi_1^f), \\
\label{equ:2} &s(\phi)=a_0^s+a_1^s\cos(\phi+\varphi_1^s),
\end{eqnarray}
where, $\phi=$mod$\left(2\pi d\alpha / p_2 + \pi, 2\pi\right) \in [-\pi, \pi]$ is the lateral offset caused by refraction angle $\alpha$; $d$ is the distance between the last two gratings and $p_2$ is the pitch of G2 grating. From Eqs. (\ref{equ:1}) and (\ref{equ:2}), regular contrasts can be expressed as  \cite{pfeiffer2006phase,wang2009quantitative},
\begin{eqnarray}
\label{equ:3} &&A_{FCA} = -\ln \left( \frac{a_0^s}{a_0^f}\right), \\
\label{equ:4} &&P_{FCA} =\frac{p_2}{2\pi d} \left(\phi_1^s-\phi_1^f \right),	\\
\label{equ:5} &&D_{FCA} = -\frac{1}{2\pi^2} \left(\frac{p_2}{d}\right)^2 \ln\left( \frac{a_1^s}{a_0^s} \Big/ \frac{a_1^f}{a_0^f} \right), 
\end{eqnarray}
\noindent where, $A, P, D$ deonting ATC, DPC, and DFC, respectively; $\phi_c = \phi_1^s - \phi_1^f$ denoting the phase shift; $\omega = 2 \pi d / p_2$; $V^s = a_1^s / a_0^s, V^f = a_1^f /a_0^f$ denoting the visibilities of the sample PSC and the flat PSC. The three parameters of PSCs ($a_0$, $a_1$ and $\phi_1$) are usually calculated by the Fourier transform, which is a fast and stable approach. However, due to the phase contrast expression in Eq. (\ref{equ:4}), FCA has an inherent phase-wrapping problem, which could be a problem in practical applications \cite{jerjen2011reduction,epple2014phase,rodgers2018quantitative}.

The DB-MMA method is based on  the well-established convolution relationship that the sample PSC can be considered as the convolution of the flat PSC with an underlying USAXS distribution  \cite{wang2009quantitative}, i.e.,
\begin{equation}
\label{equ:6} s(\phi)=f(\phi)\otimes g(\phi)
\end{equation}
where, $g(\phi)$ denoting the USAXS distribution. Then DB-MMA represents regular three contrasts as parameters of the deconvolved distribution $g(\phi)$, using a simple pattern that the ATC corresponds to the zero-order moment ($M_0$), the DPC to the first-order moment ($M_1$), and the DFC to the second-order centralized moment ($M_2$)  \cite{modregger2012imaging}, i.e.,
\begin{eqnarray}
\label{equ:7} &&A_{DB} \Rightarrow M_0(g) = \int_{-\pi}^\pi  g(\phi) d\phi, \\
\label{equ:8} &&P_{DB} \Rightarrow  M_1(g) = \frac{1}{M_0(g)}\int_{-\pi}^\pi  \phi g(\phi) d\phi, \\
&&D_{DB} \Rightarrow M_2(g) = \frac{1}{M_0(g)} \int_{-\pi}^\pi  \left( \phi - M_1(g)  \right)^2 g(\phi) d\phi \nonumber\\
\label{equ:9}&&\qquad \quad =\frac{1}{M_0(g)} \int_{-\pi}^\pi  \phi^2 g(\phi) d\phi - M_1^2(g)
\end{eqnarray}
where, the subscript ``DB'' denoting the DB-MMA method.

To deconvolve the scattering distribution $g(\phi)$, an iterative process is usually utilized, where the deconvolution method and the stopping criterion are needed to be pre-determined. DB-MMA solves the phase-wrapping problem in FCA naturally and can simultaneously obtain higher order moments like the skewness $\tilde{M}_3(g)$ \cite{modregger2012imaging} and the kurtosis ($M_4/M_2^2$) \cite{modregger2017interpretation}, which could be useful in some applications.  However, the time-consuming deconvolution process and the difficulty in determining a proper stopping criteria for the iteration make DB-MMA less-favorable for time-sensitive applications.
\subsection{Truncated analytic multi-order moment analysis}
To overcome the limitations of FCA and DB-MMA above, we will derive an analytic multi-order moment analysis as follows.  

According to the convolution relationship in Eq. (\ref{equ:6}) and the convolution property of Fourier transform, one can get
\begin{equation}
\label{equ:10} \int^{\pi}_{-\pi} e^{\pm j n \phi} s\left(\phi\right)  d\phi = \int^{\pi}_{-\pi} e^{\pm j n \phi} f\left(\phi\right)  d\phi \int^{\pi}_{-\pi} e^{\pm j n \phi} g\left(\phi\right)  d\phi,
\end{equation}
\noindent where, $n \in \mathbb{N}$ is the variable in the discrete Fourier space and $e^{\pm j n \phi}=\cos(n\phi)\pm j\sin(n\phi)$ is based on the Euler's formula.

When $n = 0$, from Eq.~(\ref{equ:10}) one can directly get the zero-order moment of the scattering distribution $g(\phi)$, i.e.,
\begin{equation}
\label{equ:11} M_0(g) =\int_{-\pi}^\pi  g(\phi) d\phi= \frac{\int^{\pi}_{-\pi} s\left(\phi\right)  d\phi}{\int^{\pi}_{-\pi} f\left(\phi\right)  d\phi}= \frac{M_0(s)}{M_0(f)}.
\end{equation}

When $n \geq 1$, from Eq.~(\ref{equ:10}) one can get the inner product of the scattering distribution $g(\phi)$ and a single $n$th-order trigonometric function, which is referred as $n$th-order trigonometric moments ($SinM_n$ or $CosM_n$) of $g(\phi)$, i.e.,
\begin{eqnarray}
\label{equ:12} &&CosM_n(g) = \int_{-\pi}^\pi \cos(n \phi) g\left(\phi\right) d\phi  \\ 
&&= \frac{CosM_n(s) \cdot CosM_n(f)+ SinM_n(s) \cdot SinM_n(f) }{CosM_n^2(f) + SinM_n^2(f) } \nonumber,  \\
\label{equ:13} &&SinM_n(g) = \int_{-\pi}^\pi \sin(n \phi) g\left(\phi\right) d\phi \\
&&= \frac{SinM_n(s) \cdot CosM_n(f)- CosM_n(s) \cdot SinM_n(f) }{CosM_n^2(f) + SinM_n^2(f) }\nonumber, 
\end{eqnarray}
\noindent where, the trigonometric moments of any function $y(\phi)$ (like $s(\phi)$ and $f(\phi)$ above) are defined as,
\begin{eqnarray}
&&CosM_n(y)= \int_{-\pi}^\pi \cos(n \phi) y\left(\phi\right) d\phi, \label{equ:14}\\
&&SinM_n(y)= \int_{-\pi}^\pi \sin(n \phi) y\left(\phi\right) d\phi. \label{equ:15}
\end{eqnarray}
From the expressions in Eqs. (\ref{equ:8}) and (\ref{equ:9}), it is observed that DB-MMA computes the integrals of the scattering distribution $g(\phi)$ with a kernel function $h(\phi)$ (i.e. $h_P(\phi)= \phi$ for the phase contrast and $h_D(\phi)= \phi^2$ for the dark-field contrast). In the range of $\phi \in [-\pi, \pi]$, one can express both kernel functions with their Fourier Series, i.e.,
\begin{eqnarray}
&&h_P(\phi) = \phi= 2 \sum_{n=1}^{\infty} \frac{(-1)^{n+1}}{n} \sin(n \phi),  \label{equ:16} \\
&&h_D(\phi) = \phi^2 = \frac{\pi^2}{3} + 4 \sum_{n=1}^{\infty} \frac{(-1)^n}{n^2} \cos(n \phi).\label{equ:17} 
\end{eqnarray}
According to Eqs. (\ref{equ:16}) and (\ref{equ:17}), the moments of the scattering distribution $g(\phi)$ in Eqs. (\ref{equ:8}) and (\ref{equ:9}) can be rewritten as the summation of trigonometric moments of $g(\phi)$ from the first-order to the infinite-order. Therefore, from Eqs. (\ref{equ:12}) and (\ref{equ:13}), the moments of $g(\phi)$ can be obtained directly by a summation of trigonometric moments of $s(\phi)$ and $f(\phi)$ from the first-order to the infinite-order, i.e.,
\begin{eqnarray}
\label{equ:18}&&M_1(g) =  \frac{2}{M_0(g)} \sum_{n=1}^\infty \left[\frac{(-1)^{n+1}}{n} \right.  \\
&&\left. \cdot \frac{SinM_n(s) \cdot CosM_n(f)- CosM_n(s) \cdot SinM_n(f) }{CosM_n^2(f) + SinM_n^2(f) }\right],   \nonumber\\
\label{equ:19} &&M_2(g) =  \frac{\pi^2}{3} - M_1^2(g) + \frac{4}{M_0(g)}  \sum_{n=1}^\infty \left[ \frac{(-1)^n}{n^2} \right. \\
&&\left.  \cdot   \frac{CosM_n(s) \cdot CosM_n(f)+ SinM_n(s) \cdot SinM_n(f) }{CosM_n^2(f) + SinM_n^2(f) } \right]. \nonumber 
\end{eqnarray}
Eqs. (\ref{equ:11}), (\ref{equ:18}) and (\ref{equ:19}) constitute a new multi-order moment analysis, which is totally analytic. However,  Eqs. (\ref{equ:18}) and (\ref{equ:19}) are just theoretical formulas and cannot be realized in practice  due to the summation of infinite order terms. As a result, we will propose a more practical form below. 

According to the trigonometric orthogonality, it's known that,
\begin{eqnarray}
&&\int_{-\pi}^ {\pi} \cos \left(n\phi \right) \left[a_0 + a_1\cos \left(\phi + \varphi \right)\right] = 0, \label{equ:20}\\
&&\int_{-\pi}^ {\pi} \sin \left(n\phi \right) \left[a_0 + a_1 \cos \left(\phi + \varphi \right)\right] = 0, \label{equ:21} 
\end{eqnarray}
\noindent when $\forall n \geq 2$ and $n \in \mathbb{N}, \forall \varphi \in \mathbb{R}$. It is worth noting that all the second or higher order trigonometric moments of PSCs including $CosM_n(s)$, $SinM_n(s)$, $CosM_n(f)$ and $SinM_n(f)$ are very close to zero because PSCs are approximate to the cosine models in Eqs. (\ref{equ:1}) and (\ref{equ:2}). Therefore, the $n$th-order trigonometric moments of $g(\phi)$ calculated by Eqs. (\ref{equ:18}) and (\ref{equ:19}) could be numerically unstable when $n \geq 2$.
For computation stability, one can obtain a truncated form of the proposed analytic analysis above as,
\begin{eqnarray}
\label{equ:22} &&M_1(g) \simeq 2\frac{SinM_1(s) CosM_1(f)- CosM_1(s) SinM_1(f) }{M_0(g)\cdot\left[CosM_1^2(f) + SinM_1^2(f) \right]}, \nonumber\\ 
&&\\
\label{equ:23} &&M_2(g) \simeq  \frac{\pi^2}{3} - M_1^2(g) - \frac{4}{M_0(g)}\nonumber \\
&&\cdot   \frac{CosM_1(s) CosM_1(f)+ SinM_1(s) SinM_1(f) }{CosM_1^2(f) + SinM_1^2(f) }. 
\end{eqnarray}
With Eqs. (\ref{equ:11}), (\ref{equ:22}) and (\ref{equ:23}) above, we thereby establish a stable, efficient and analytic information retrieval method (i.e., TA-MMA) that can obtain multiple contrasts directly from raw PSCs, without the deconvolution process in DB-MMA, leading to hundreds of times faster in computation than the original DB-MMA. It is worth noting that $M_1(g)$ and $M_2(g)$ in TA-MMA are zero-order and first-order components  in Fourier Series of corresponding moments in MMA, which could be a good approximation in practice as demonstrated in the ``RESULTS AND DISCUSSIONS'' section.
\subsection{Physical interpretations of TA-MMA and its hybrid-field contrast}
Based on the analyses above, we know that TA-MMA can be a good approximation of MMA, and it gets rid of the complicated deconvolution in DB-MMA, making it comparable with FCA in terms of the computational efficiency. Meanwhile, TA-MMA can serve as an important bridge to connect FCA and MMA, which are totally different forms of information retrieval algorithms. Following the tricks by Li et al. \cite{li2019comparative}, if we apply the cosine models of FCA in Eqs. (\ref{equ:1}) and (\ref{equ:2}) to the analytic expressions of TA-MMA in Eqs. (\ref{equ:11}) (\ref{equ:22}) (\ref{equ:23}), the contrasts retrieved by TA-MMA with parameters in cosine models can be expressed as, i.e.,
\begin{eqnarray}
\label{equ:24}&&A_{TA} =-\ln M_0(g)= - \ln \frac{a_0^s}{a_0^f}, \\
\label{equ:25} &&P_{TA} = \frac{1}{\omega}M_1(g) = -  \frac{2}{\omega}\frac{V^s}{V^f} \sin (\phi_c), \\
\label{equ:26} &&D_{TA} =\frac{1}{\omega^2} M_2(g) \\
&&= \frac{\pi^2}{3} \frac{1}{\omega^2}-  \frac{4}{\omega^2} \left(\frac{V^s}{V^f}  \right)^2 \sin^2 \phi_c -  \frac{4}{\omega^2} \frac{V^s}{V^f}\cos \phi_c.\nonumber 
\end{eqnarray}
\noindent where, the subscript ``TA'' denoting the TA-MMA method.
\begin{figure*}[htbp]
\includegraphics[width=\textwidth]{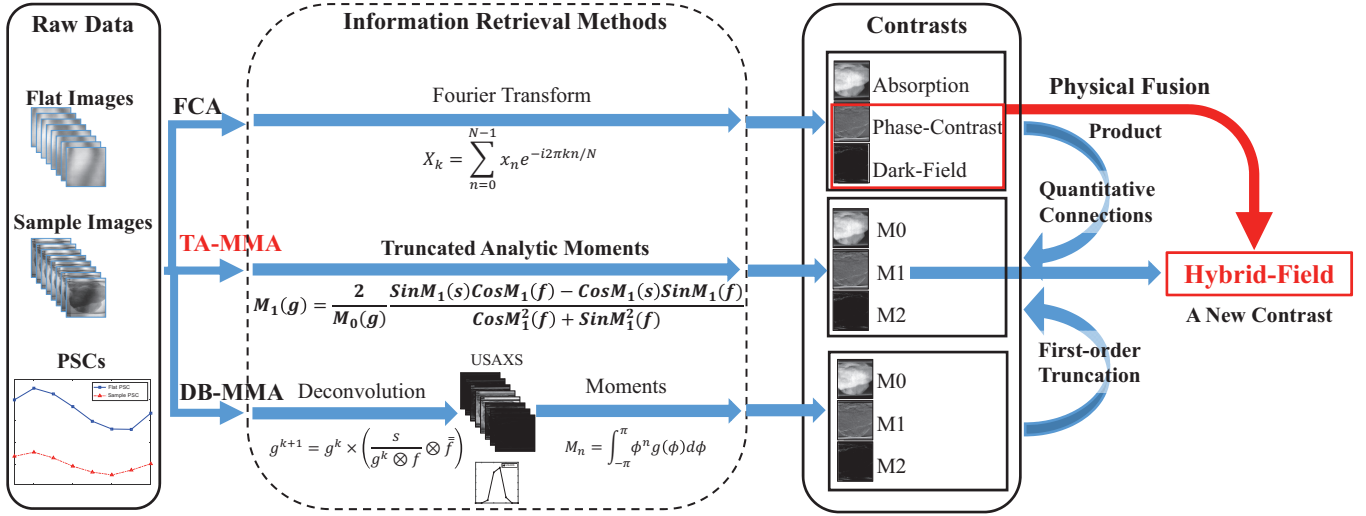}%
\caption{The schematic diagrams of the conventional FCA, DB-MMA and our proposed TA-MMA.}%
\label{fig:diagram}
\end{figure*}

From Eqs. (\ref{equ:24}) - (\ref{equ:26}), we immediately establish a quantitative connection between FCA and MMA. Consistent with the intrinsic relationship revealed by Li et al. \cite{li2019comparative}, both the first-order moment and the second-order moment retrieved by MMA are combinations of the original DPC and DFC retrieved by FCA. Furthermore, as the DPC in Eq. (\ref{equ:4}) by FCA is represented by the phase shift $\phi_c=\phi_1^s-\phi_1^f$ and the DFC in Eq. (\ref{equ:5}) by FCA is represented by the visibility ratio $V^s/V^f$, it is clearer to see that the first-order moment retrieved in Eq. (\ref{equ:25}) by TA-MMA is actually the product of the DPC and the DFC retrieved by FCA, under the small phase shift assumption (i.e., $\sin(\phi_c)\approx\phi_c$), which can be justified in biological samples like breast tissues. As shown in Fig. \ref{fig:diagram}, the schematic diagrams of FCA, DB-MMA and our proposed TA-MMA reflect the advantages of TA-MMA both in higher computation efficiency and clearer physical interpretations.

Moreover, if we take the logarithm of both sides' absolute values in Eq. (\ref{equ:25}), the equation turns to be,
\begin{equation}
\label{equ:27} \ln  |P_{TA-MMA}| = \ln \left[\frac{V^s}{V^f}\right] + \ln [|\sin(\phi_c)|] +\ln [\frac{1}{\omega}].
\end{equation}
From Eq. (\ref{equ:27}), it is seen that the the contrast caused by visibility ratios can be dominant component when the phase shift is low. In these cases, the logarithmic absolute phase contrast retrieved by TA-MMA is almost determined by the logarithmic visibility ratio, which is just the DFC retrieved by FCA in Eq. (\ref{equ:5}).

Consequently,  the first-order moment retrieved by TA-MMA in Eq. (\ref{equ:22}) can be used as a new physical parameter fusing the original DPC and the DFC retrieved by FCA. It is a hybrid-field contrast essentially, which can combine the high-sensitivity for low-Z materials of the DPC and the sub-pixel scattering-sensitivity of the DFC together. Such a hybrid-field contrast can offer a more compact image representation for radiologists, arguably the first physical fusion image in GPCI. 
\subsection{Experimental Setups and Evaluations}
In order to validate our analyses above, experiments were carried out on a integrated GPCI platform with a conventional x-ray tube in Tsinghua University, Beijing, China, whose details can be found in Ref.  \cite{li2019quantitative}. 

In this work, we utilized two groups of gratings with different imaging parameters for two GPCI systems, including a coherent Talbot-Lau interferometry and an incoherent geometrical-projection system. The system parameters and experimental specimens are listed in Table. \ref{tab:table0}.

\begin{table}[htbp]
\begin{center}
\caption{\label{tab:table0} The parameters of experimental GPCI imaging systems used in our work.}
\begin{tabular}{ccc}
\hline \hline
System Type&Talbot-Lau&geometrical-projection\\ \hline
Source&\multicolumn{2}{c}{Comet MXR-160HP/11}\\
Tube voltage&35 kV& 35kV\\
Tube current&35 mA&35mA\\
Exposure time/step& 800 ms & 800ms\\
Detector&\multicolumn{2}{c}{Dexela 1512}\\
Pixel Size&75 $\mu$m&75 $\mu$m\\
G0 pitch&16.8 $\mu$m&42 $\mu$m\\
G1 pitch&4.2 $\mu$m&6 $\mu$m\\
G1 type&$\pi$ phase &absorption\\
G2 pitch&2.4 $\mu$m&7 $\mu$m\\
G1 to G2 distance&26 cm&17 cm\\
Number of Steps&10&10\\
\multirow{2}{*}{Specimens}&a ex-vivo rat& a rat bone joint\\
&a breast tissue&\\
\hline\hline
\end{tabular}
\end{center}
\end{table}

Specially, for the breast tissue specimen, it was scanned with $1^\circ$ increments over $360^\circ$ for computed tomography (CT) reconstruction using the conventional filtered backprojection algorithm.  Additionally, in our implementation of the DB-MMA method, the scattering distribution was retrieved by 200 iterations of the Lucy-Richardson deconvolution.

In order to quantitatively evaluate different methods, an index called structural similarity (SSIM)  \cite{wang2004image} is utilized to compare the retrieved images by different information retrieval methods, which is defined as
\begin{equation}
\label{equ:28} SSIM(u,v) = \frac{(2\mu_u \mu_v+C_1)(2\sigma_{uv}+C_2)}{(\mu_u^2+\mu_v^2+C_1)(\sigma_u^2+\sigma_v^2+C_2)},
\end{equation}
\noindent where, $\mu_u$, $\mu_v$, $\sigma_u$, $\sigma_v$, and $\sigma_{uv}$ are the local means, standard deviations, and cross-covariance for images u and v, respectively. The resultant SSIM index is a decimal value between -1 and 1, and the value 1 only occurs in the case of two identical sets of data. In this work, all images are normalized to the range of $[0, 1]$, and it is set that $C_1 = 1.0 \times 10^{-4}, C_2 = 9.0 \times 10^{-4}$ and $C_3 = 4.5 \times 10^{-4}$. 

Besides, the contrast-to-noise ratio (CNR)  \cite{sztrokay2013assessment}, defined as $|S_A-S_B|/\sigma_0$, is also calculated as another quantitative evaluation index, where $S_A$ and $S_B$ are the measurements between two defined structures and $\sigma_0$ is the standard deviation of image values in a background region.

\section{\label{sec:results} Results and Discussions}
\begin{figure}[htbp]
	\includegraphics[width=0.992\columnwidth]{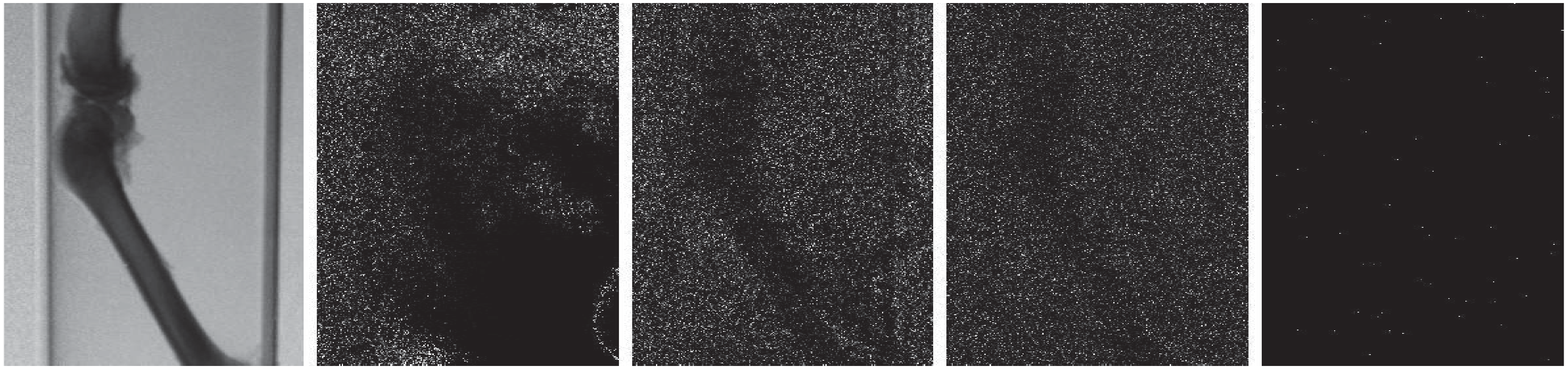}\\%
	\includegraphics[width=1.0\columnwidth]{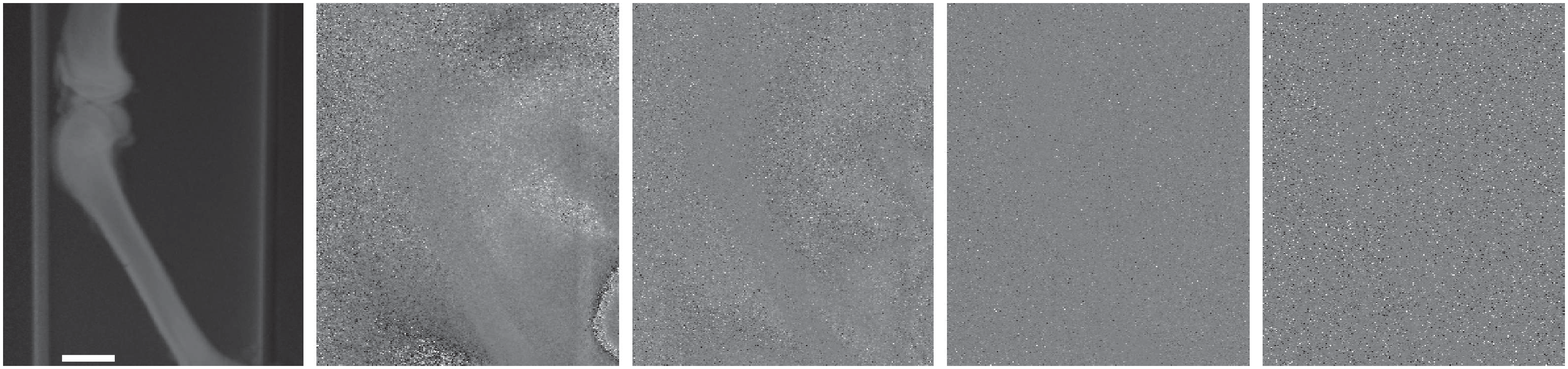}\\
	\caption{\label{fig:FS_g}Trigonometric moments of $M_1(g)$ (the first row) and $M_2(g)$ (the second row) in the theoretical analytic analysis from the first-order to the fifth-order when using the rat bone specimen, corresponding to Eqs. (\ref{equ:18}) and(\ref{equ:19}) respectively.  Scale bar here is 6 mm.}%
\end{figure}

In order to demonstrate the rationality of the first-order truncation in TA-MMA, we should first take a look at the core trigonometric moments of $M_1(g)$ and $M_2(g)$ in the proposed theoretical analytic analysis, corresponding to Eqs. (\ref{equ:18}) and (\ref{equ:19}). The trigonometric moments from the first-order to the fifth-order of the rat bone specimen are shown in Fig. \ref{fig:FS_g}, and the $i$-order components from the sixth-order to the ninth-order (the maximum order limited by the number of phase-stepping process) is the same with the $(10-i)$-order components due to the symmetry of trigonometric functions. It is seen that only the first-order trigonometric moments of both moments show detailed structures of the rat bone joint specimen. So it is reasonable and reliable to use the first-order truncation of the two moments in Eqs. (\ref{equ:18}) and (\ref{equ:19}) for approximation.

\begin{figure}[htbp]
	\includegraphics[width=0.9\columnwidth]{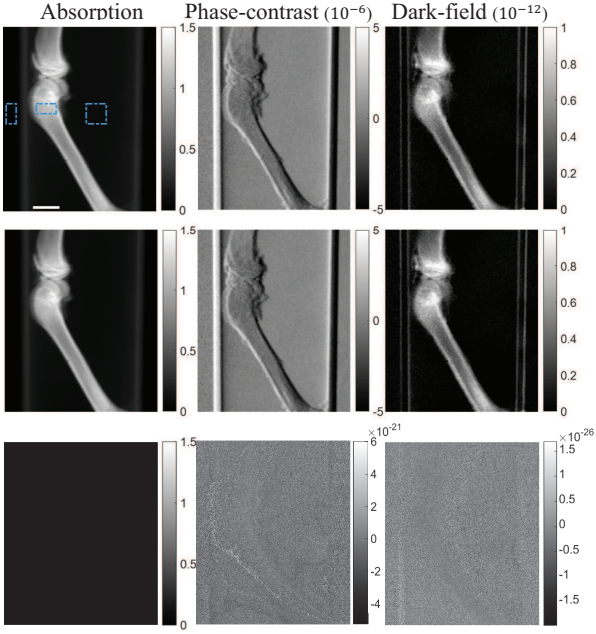}%
	\caption{\label{fig:Comp1}The regular contrasts retrieved by TA-MMA (top), the indirect computation from FCA (middle) and their difference images (bottom) using the rat bone specimen. The indirect computation from FCA means to apply contrasts retrieved by FCA to Eqs. (\ref{equ:24}) - (\ref{equ:26}) and obtain the contrasts retrieved by TA-MMA indirectly. It is seen that both approaches deliver practically identical results and their differences are significantly lower than their values. Scale bar here is 6 mm. }%
\end{figure}

According to Eqs. (\ref{equ:24}) - (\ref{equ:26}), we can indirectly compute the contrasts retrieved by TA-MMA with the contrasts retrieved by FCA in Eqs. (\ref{equ:3}) - (\ref{equ:5}). In order to validate the quantitative connections in Eqs. (\ref{equ:24}) - (\ref{equ:26}), we compared the contrasts retrieved by TA-MMA and the indirect computation from FCA when using the rat bone specimen. As shown in Fig. \ref{fig:Comp1}, it is seen that the retrieved contrasts by both methods are almost the same. The quantitative performance comparison of both approaches is listed in Table \ref{tab:Comp1}. From Table \ref{tab:Comp1}, it is found that SSIM values between TA-MMA and the indirect computation from FCA are all 1.000, and their corresponding CNRs are the same, which validates our analysis of the quantitative connections between FCA and MMA in Eqs. (\ref{equ:24}) - (\ref{equ:26}).\\

\begin{table}[htbp]
	\centering
	\caption{\label{tab:Comp1}Comparisons of TA-MMA and FCA-MMA shown in Fig. \ref{fig:Comp1}. The CNR values are computed from two structure areas and a background areas indicated by blue rectangles in Fig. \ref{fig:Comp1}.}
	\begin{tabular}{llccc}
	\hline\hline
	\multicolumn{2}{c}{Contrast}& ATC & DPC & DFC  \\
	\hline
	\multicolumn{2}{c}{SSIM}  & \textbf{1.00} & \textbf{1.00} & \textbf{1.00} \\
	\hline
	\multirow{2}{*}{CNR}&TA-MMA &279.81&9.21&10.81\\
	&FCA-MMA &279.81&9.21&10.81\\
	\hline\hline
	\end{tabular}
\end{table}

\begin{figure}[htbp]
	\includegraphics[width=1.0\columnwidth]{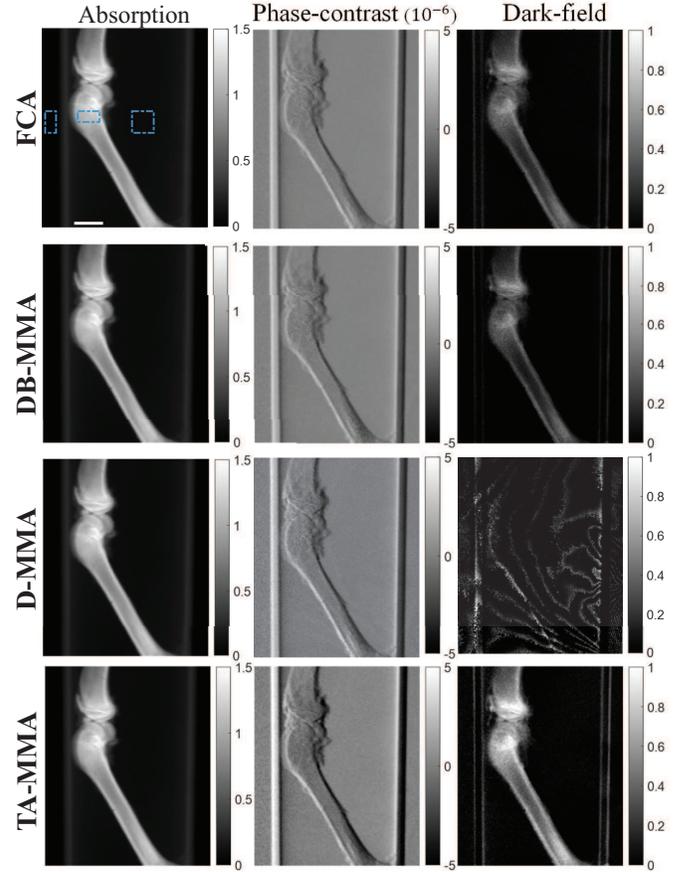}%
	\caption{\label{fig:Comp2}Comparisons of the three modalities retrieved by FCA, DB-MMA, D-MMA and TA-MMA for the rat bone specimen. Scale bar here is 6 mm.}%
\end{figure}

To compare contrasts retrieved by TA-MMA with those by FCA, DB-MMA and D-MMA, contrasts of the simple rat bone specimen are shown in Fig. \ref{fig:Comp2}. It is seen that the ATCs and DPCs retrieved by various methods remain the same. D-MMA does not work very well for the DFC even if after some data pre-processing according to Zhu et. al.  \cite{zhu2019direct}, which demonstrates D-MMA is inapplicable in GPCI in this case. In general, most detailed structures in contrasts by the other three methods (FCA, DB-MMA and TA-MMA) can be observed, while DPC values and DFC values retrieved by TA-MMA are higher than those by FCA and DB-MMA, in particular for the DFC signals. 

\begin{figure}[htbp]
	\includegraphics[width=1.0\columnwidth]{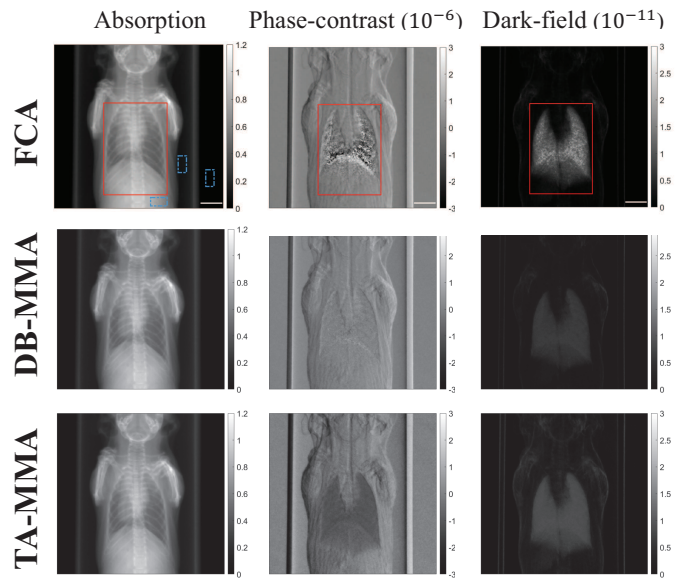}%
	\caption{\label{fig:Comp3}Comparisons of the three modalities retrieved by FCA, DB-MMA and TA-MMA for the ex-vivo rat specimen. Scale bar here is 6 mm.}%
\end{figure}
 
\begin{figure}[htbp]
	\includegraphics[width=1.0\columnwidth]{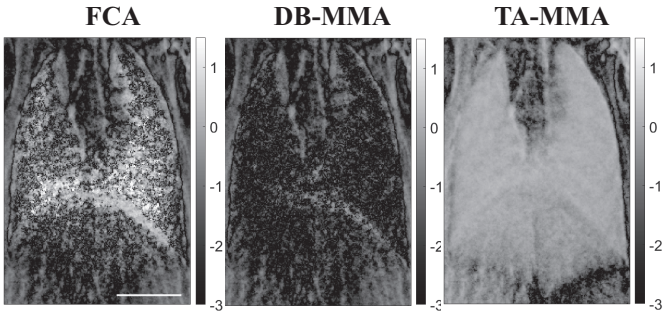}%
	\caption{\label{fig:Comp4}The logarithmic absolute DPC (defined in Eq. \ref{equ:27}) computed by FCA, DB-MMA and TA-MMA in the lung area of the rat specimen. It's seen that the logarithmic absolute DPC computed by TA-MMA is very similar with the DFC in Fig. \ref{fig:Comp3}. Scale bar here is 6 mm.}%
\end{figure}

To further evaluate the characteristics of the different information retrieval methods, multiple contrasts of a typical rat lung specimen are shown in Fig. \ref{fig:Comp3}. As expected, one can see obvious phase-wrapping artifacts in the DPC image computed by FCA, while both moment analysis methods avoid the problem very well. Moreover, the DFC images by all methods show the lower lung clearly, which is blocked by the heart in the ATC images (as indicated in the solid red rectangular regions in Fig. \ref{fig:Comp3}). 

\begin{figure}[htbp]
	\includegraphics[width=1.0\columnwidth]{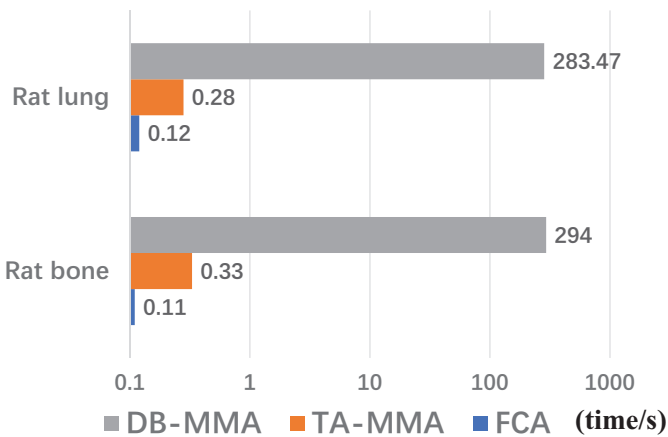}%
	\caption{\label{fig:time}The computation time of FCA, DB-MMA and TA-MMA in both the rat bone and ex-vivo rat lung specimen.}%
\end{figure}

\begin{table*}[htbp]
	\centering
	\caption{\label{tab:Comp23}Performance comparisons of FCA, DB-MMA and TA-MMA for regular contrasts shown in Fig. \ref{fig:Comp2} and Fig. \ref{fig:Comp3}. The CNR values are computed from two structure areas and a background areas indicated by blue rectangles in Fig. \ref{fig:Comp2} and Fig. \ref{fig:Comp3}.}
	\begin{tabular}{llcccccc}
	\hline\hline
		&Method& \multicolumn{2}{c}{FCA vs DB-MMA} &\multicolumn{2}{c}{DB-MMA vs TA-MMA} & \multicolumn{2}{c}{FCA vs TA-MMA}  \\
		\cline{3-4} \cline{5-6} \cline{7-8}
		&Specimen&Rat Bone&Rat Lung&Rat Bone&Rat Lung&Rat Bone&Rat Lung\\
		\hline
		\multirow{3}{*}{SSIM}&ATC & 1.00&1.00 &  1.00&1.00 & 1.00&1.00\\
		&DPC & 0.983& 0.950 & 0.906&0.950 &0.910&0.910\\
		&DFC &0.960&0.942 &0.839&0.922 &0.889&0.913\\
		\hline
		&Method& \multicolumn{2}{c}{FCA} &\multicolumn{2}{c}{DB-MMA} & \multicolumn{2}{c}{TA-MMA}  \\
		\cline{3-4} \cline{5-6} \cline{7-8}
		&Specimen&Rat Bone&Rat Lung&Rat Bone&Rat Lung&Rat Bone&Rat Lung\\
		\hline
		\multirow{3}{*}{CNR}&ATC& 279.81&198.28&279.81&198.28&279.81&198.28\\
		&DPC & 1.40&10.22&2.67&10.21&9.21&5.79\\
		&DFC &14.75&6.49 &40.79&8.72&10.81&4.07\\
		\hline\hline
	\end{tabular}
\end{table*}

Meanwhile, the blocked lower lung can also be found in the hybrid-field contrast (i.e., the original DPC) image by TA-MMA, which suggests the hybrid-field contrast computed by TA-MMA contains partial dark-field information indeed. As demonstrated in Eq. (\ref{equ:27}), under small phase shift conditions, the logarithmic absolute hybrid-field contrast computed by TA-MMA can be approximated to the original DFC. The alveoli in the lung region are mainly composed of air, which is a perfect specimen with small phase shifts to validate our analyses. As shown in Fig. \ref{fig:Comp4}, it is seen that the logarithmic absolute hybrid-field contrast computed by TA-MMA is very similar to the DFC, while the same process to the images by FCA and DB-MMA have very little information. These comparison results demonstrate the advantages of the new hybrid-field contrast, i.e., the first-order moment computed by TA-MMA may be able to represent both the original DPC and DFC in some practical applications.

\begin{figure}[htbp]
	\includegraphics[width=1.0\columnwidth]{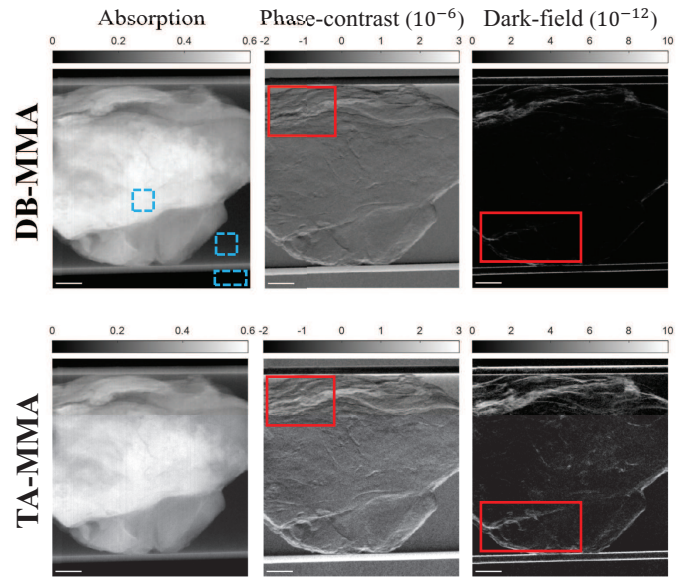}%
	\caption{\label{fig:Breast1}Comparisons between the three modalities retrieved by DB-MMA and TA-MMA for the breast tissue specimen. Scale bar here is 6 mm.}%
\end{figure}

\begin{figure}[htbp]
	\includegraphics[width=1.0\columnwidth]{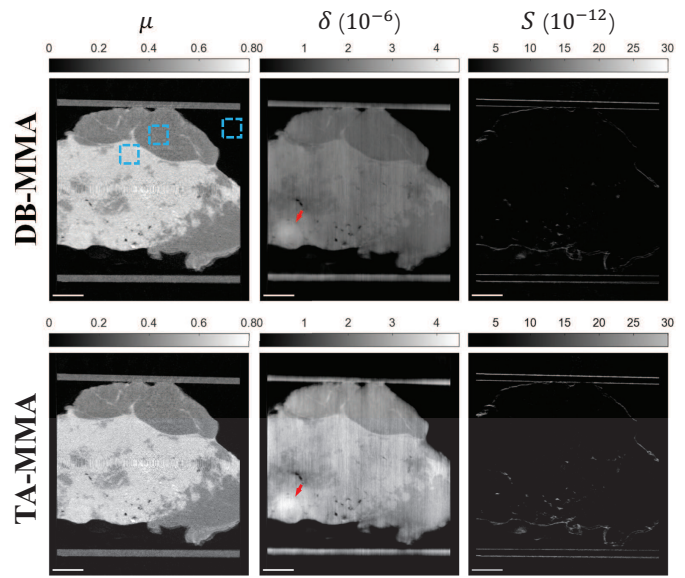}%
	\caption{\label{fig:Breast2}Comparisons between the three reconstructed modalities retrieved by DB-MMA and TA-MMA for the breast tissue specimen. Scale bar here is 6 mm.}%
\end{figure}

Quantitative evaluations for the rat bone and rat lung specimens were performed and summarized in Table \ref{tab:Comp23}. In terms of SSIM, all three methods accord well with each other as their SSIMs are all quite close to 1. In terms of CNR, it is seen that all three methods have similar values, while the CNR value of TA-MMA is a little lower. These differences may result from the smoothing effect by the iterations of deconvolution in DB-MMA as demonstrated in Ref.  \cite{weber2013increasing}. If needed, one can improve the noise performance of TA-MMA by some post-processing denoising algorithms, which is quite typical in medical applications. Besides, as shown in Fig. \ref{fig:time}, the computation time of TA-MMA in both specimens is much less than that of DB-MMA, which demonstrates the high efficiency of TA-MMA. In general, our proposed TA-MMA is a reliable and efficient information retrieval method in GPCI.

Finally, we conducted a simple preclinical test for the diagnosis of breast tumors by applying the GPCI technology to a woman's breast tissue specimen. As shown in Fig. \ref{fig:Breast1} and Fig. \ref{fig:Breast2}, we computed both projection contrasts and CT reconstruction images by DB-MMA and TA-MMA. Quantitative evaluations for the breast specimen are summarized in Table \ref{tab:Breast}. As expected, there are good agreements between DB-MMA and TA-MMA on the ATC projections and CT images. One  can see TA-MMA offers more detailed structures on the DPC and DFC (as highlighted in the solid red rectangular regions in Fig. \ref{fig:Breast1}), and the associated CT images are of higher signals than DB-MMA. What's more, it was noting that for both DB-MMA and TA-MMA, the tumor lesions are clearly observed on the DPC CT images (indicated by the red arrows in Fig. \ref{fig:Breast2}) but not on ATC CT images, which demonstrates the potential of GPCI in the diagnosis of breast tumors. In addition, one can also see in Table \ref{tab:Breast} that DPC images and DFC images by TA-MMA in this specimen are even of a little better CNR values compared with those by DB-MMA.

\begin{table}[htbp]
	\centering
	\caption{\label{tab:Breast}Performance comparisons of TA-MMA and DB-MMA shown in Fig. \ref{fig:Breast1} and Fig. \ref{fig:Breast2}.}
	\begin{tabular}{lcccccc}
	\hline\hline
	\multicolumn{3}{c}{Contrast}& ATC & DPC & DFC   \\
	\hline
	\multirow{2}{*}{SSIM}&\multicolumn{2}{c}{Projection}  & 1.00 & 0.87 & 0.64 \\
	&\multicolumn{2}{c}{CT}&1.00&0.94&0.72\\
	\hline
	\multirow{4}{*}{CNR}&\multirow{2}{*}{Projection}&DB-MMA &101.17&1.90&0.99\\
	& &TA-MMA &101.17&1.97&1.42\\
	&\multirow{2}{*}{CT}&DB-MMA &14.75&5.78&0.52\\
	& &TA-MMA &14.75&6.19&0.98\\
	\hline\hline
	\end{tabular}
\end{table}

\section{\label{sec:conc}Conclusion}
In this paper, we present an analytic form of multi-order  moments  analysis in theory that can retrieve multiple contrasts directly from phase-stepping images with no scattering distribution involved. For practical implementation, a first-order truncation of the proposed theoretical analysis is adopted and it is totally analytic, efficient and stable, leading to hundreds of times faster in computation than the original deconvolution-based multi-order moment analysis. More importantly, our proposed truncated analytic moment analysis is proved to establish a quantitative connection between Fourier components analysis and multi-order moment analysis, i.e., the first-order moment computed by our proposed method is essentially the product of the phase contrast and the dark-field contrast retrieved by Fourier components analysis, providing hybrid-field contrast as a new physical parameter. The hybrid-field contrast fuses the original phase contrast and dark-field contrast in a straightforward manner, which may be the first physical fusion contrast in GPCI to our knowledge and may have a potential to be directly used in practical applications.

\ifCLASSOPTIONcaptionsoff
  \newpage
\fi

\end{document}